\magnification 1200

\rightline{KCL-MTH-10-15}

\vskip .5cm
\centerline {\bfÊ $\bf E_{11}$ and Supersymmetry }

\vskip 1cm
\centerline{Duncan Steele and Peter West}
\centerline{Department of Mathematics}
\centerline{King's College, London WC2R 2LS, UK}

\vskip .5 cm
\noindent
We introduce fermions into the $E_{11}$ non-linear realisation. We show,
at low levels, that the commutators of the Cartan involution invariant
subalgebra of $E_{11}$ with the known supersymmetry transformations of
eleven dimensional supergravity lead to symmetries of the theory
indicating the consistency of supersymmetry and $E_{11}$.Ê

\vfillÊ
\eject

\medskip
{\bf 1. Introduction}
\medskip
The first paper which conjectured $E_{11}$ symmetry [1]Ê only
considered the bosonic sectors of the theories under consideration and 
the same is true for  all subsequent papers. However, there have been
a number of results which follow from
$E_{11}$ symmetry which have traditionally been,  or have subsequently
 been, shown to follow from  supersymmetry.  Two such examples are
the two and five form central charges in the $l_1$ representation of 
$E_{11}$, which is conjectured to contain all  brane
charges,  [2] and the representations carried by form fields which imply
the classification ofÊ gauged supergravities [3,4]. A brief account of
some  of the evidence for an underlying $E_{11}$ symmetry of strings and
branes is summarisedÊ in the first seven pages of [5]. 
\par
The dimensionally reduced maximal supergravitiesÊ contain 
non-linear realisations that encode the scalar fields. In fact this
statement is true for all supergravity theories which possess scalars in
their supergravity multiplet. The prototype example  is the maximal
supergravity theory in four dimensions which possesses an 
$E_7$ non-linear realisation with  local subgroup SU(8) [6].  The
other fields in the supergravity multiplet transform as matter
representations of the non-linear realisation, that is under the local
subgroup. This   includes the  fermions.  
\par
The local subalgebra adopted in the non-linear realisation of $E_{11}$
is theÊ Cartan involution invariant subalgebra denoted $K(E_{11})$.Ê
The commutation relations of this latter algebra wereÊ given at low
levels in [2] and it was found that theÊ generators could be
represented at low levels by the eleven dimensional $\gamma$-matrices.
As such $K(E_{11})$ possesses, at low levels,  a 32 component spinor
representation that might be used as the supersymmetry parameter [2].Ê 
\par
It has also  been proposed [7] that non-linearly realised $E_{10}$ is  
a symmetry of maximal eleven dimensional supergravity.Ê This is a
subalgebra of $E_{11}$, but it differs from the earlier proposal [1] in
the way it incorporates space-time; the fields are taken to depend only
on time and the spatial derivatives of the fields are proposed to occur
at higher level in $E_{10}$. Fermions have been incorporated in
the $E_{10}$ non-linear realisation [8-11]. Following the pattern  found
in supergravity theories in lower dimensions these authorsÊ took the
fermions to belong to linear representations of
$K(E_{10})$.  They found that there exists at low levelsÊ a
 representation which is a vector spinor of the ten dimensional
Lorentz group which can be identified with the gravitino and that this is
a representation at all levels, albeit an unfaithful one. The previously
found [2]Ê thirty twoÊ component unfaithful representation, which is a
spinor of the ten dimensional Lorentz group,  was used asÊ the
supersymmetry parameter.Ê
\par
In this paper we follow a similar path to incorporate the fermions into
the $E_{11}$ non-linear realisation. The contents of this paper are as
follows. In section two we summarise the algebra of $K(E_{11})$,Ê in
section three we compute the $K(E_{11})$Ê transformations of the fields
at low levels, in section four we will find a unfaithful representation
of $K(E_{11})$ which can be identified with the gravitino, in section
five we compute the commutators of the low level fields between the known
supersymmetry transformations and their previously found $K(E_{11})$
transformations and show that they are consistent in that they lead to
known symmetries of the theory. Ê
\par
As very briefly indicated in reference [11] some calculations
incorporating fermions,Ê which are unpublished,Ê have been carried out in
the $E_{11}$ context by these authors. Ê

\medskip
{\bf 2. The Cartan Involution invariant subgroup of $E_{11}$}
\medskip
In this section we summarise the commutation relations of
$E_{11}$ generators and those ofÊ the Cartan involution invariant subgroup
of $E_{11}$, denoted $K(E_{11})$ . The Dynkin diagram of $E_{11}$ is
given by Ê$$ \matrix { Ê& & & & & & & & & & & & & &\circ&11&\cr Ê& & & &
& & & & & & & & & &|& &\cr
Ê\circ&-&\circ&-&\circ&-&\circ&-&\circ&-&\circ&-&\circ&-&\circ&-&\circ&-&\circ
\cr Ê1& &2& &3& &4& &5& &6& &7& &8& &9& &10\cr } $$
\centerline{Figure 1.Ê The Dynkin diagram of $E_{11}$}

Deleting node eleven we find the algebra  GL(11), which corresponds in
the non-linear realisation to eleven dimensional gravity. As such it is
natural to decompose the adjoint representation of $E_{11}$ in terms of
GL(11)  which consists of SL(11) and the remaining generator of the
Cartan subalgebra. We  denoted these generators 
by $K^a{}_b, \ a,b=1,\ldots , 11$Ê and they obey the commutatorsÊ
$$
Ê[K^a{}_b, K^c{}_d] = \delta^c_b K^a{}_d - \delta^a_d K^c{}_b
\eqno(2.1)$$

All generators of a Kac-Moody algebra are formed from multiple
commutators of the Chevalley generators. The level of a generator is
defined to be the number of times the Chevalley generator $E_{11}$ (not
to be mistaken with the symbol for the algebra itself)  occurs for positive
root generators, or minus the number of times
$F_{11}$ appears for negative root generators. The results of the
decomposition can be classified by this level [12,7]. The positive root
generators at level one and two respectively are given byÊ[1]
$$
ÊR^{a_1 \dots a_3}, R^{a_1 \dots a_6}
\eqno(2.2)$$
while Ê
$$
ÊR_{a_1 \dots a_3}, R_{a_1 \dots a_6}
\eqno(2.3)$$
are the  negative root generators at levels -1 and -2 respectively.

The commutation relations of the positive root generators with  GL(11)  
are 
$$
Ê[K^b{}_c, R^{a_1 \dots a_3}]
Ê = 3 \delta^{[a_1}_c R^{|b| a_2 a_3] },\ [K^b{}_c, R^{a_1 \dots a_6}] 
= 6 \delta^{[ a_1}_c R^{|b|a_2 \dots a_6] }
\eqno(2.4)$$
While the commutators of the negative root generator with those of
 GL(11)  are given by
$$
Ê[K^b{}_c, R_{a_1 a_2 a_3}] = - 3 \delta_{[a_1}^b R_{|c|a_2 a_3 ]},\quadÊ
[K^b{}_c, R_{a_1 \dots a_6}] = - 6 \delta_{[a_1}^b R_{|c|a_2 \dots
a_6 ]}
\eqno(2.5)$$
\par
Generators at level two, or minus two, can be found as the commutator
of two level one, or minus one, generators. Ê
$$
Ê[ R^{a_1 \dots a_3}, R^{a_4 \dots a_6}] = 2 R^{a_1 \dots a_6},\quadÊ
Ê[R_{a_1 \dots a_3}, R_{a_4 \dots a_6}] = 2 R_{a_1 \dots a_6}
\eqno(2.6)$$
In these equations we have chosen the normalisation of these generators.

Finally, the commutators between the positive and negative root generators
at levels one and two are [1]Ê
$$ \matrix {
Ê[R^{a_1 \dots a_3}, R_{b_1 \dots b_3}]&=&
Ê 18 \delta^{[a_1a_2}_{[b_1 b_2} K^{a_3]}{}_{b_3]}
Ê - 2 \delta^{a_1\dots a_3}_{b_1\dots b_3}\left(\sum_b K^b{}_b\right) \cr
Ê[R^{a_1 \dots a_3}, R_{b_1 \dots b_6}] &=&
Ê {5! \over 2} \delta^{a_1 \dots a_3}_{[b_1 \dots b_3} R_{b_4 \dots b_6]} \cr 
Ê[R^{a_1 \dots a_6}, R_{b_1 \dots b_6}] Ê &=& -5! \left( 9 \delta^{[a_1
\dots a_5}_{[b_1 \dots b_5} K^{a_6]}{}_{b_6]} Ê - \delta^{[a_1 \dots
a_6]}_{[b_1 \dots b_6]} \sum_c K^c{}_c \right) 
\cr 
Ê[R_{b_1 \dots b_3}, R^{a_1 \dots a_6}]
Ê &=& {5! \over 2} \delta^{[a_1 \dots a_3}_{b_1 \dots b_3} 
R^{a_4 \dots a_6]} \cr }
 \eqno(2.6)$$
\par
The usually adopted  Cartan involution of a Lie algebra is defined on the
Chevalley generators as
$$
ÊE_a \rightarrow -F_a, F_a \rightarrow -E_a, H_a \rightarrow -H_a
\eqno(2.7)$$
The effect on the generators used above is 
$$
K^a{}_b \rightarrow -K^b{}_a, R^{a_1 \dots a_3} \rightarrow -
R_{a_1 \dots a_3}, R^{a_1 \dots a_6} \rightarrow R_{a_1 \dots a_6}
\eqno(2.8)$$

The Cartan involution invariant subalgebra $K(E_{11})$ is
generated  by the invariant combination of the Chevalley generators given
by 
$$
ÊS_a = E_a - F_a
\eqno(2.9)$$

A basis for the Cartan involution invariant subalgebra  K($E_{11}$) is
given, up to and including level 2,  by  [2]
$$ \matrix {
ÊJ^{ab} &=& K^a{}_c\eta^{cb}-  K^b{}_c\eta^{ca} \cr
ÊS_{a_1 \dots a_3} &=& R^{b_1 \dots b_3} \eta_{b_1 a_1} 
\dots \eta_{b_3 a_3} - R_{a_1 \dots a_3} \cr
ÊS_{a_1 \dots a_6} &=& R^{b_1 \dots b_6} \eta_{b_1 a_1} 
\dots \eta _{b_6 a_6} + R_{a_1 \dots a_6} \cr
}Ê
\eqno(2.10)$$
The $J^{ab}$s generate the Lorentz algebra. In fact, these generators are
not the generators which are invariant under the Cartan involution
of equation (2.7), but under the modified Cartan involution given by
$E_a\rightarrow -\eta_{aa} F_a,\  F_a\rightarrow -\eta_{aa}  E_a$ and $
H_a\rightarrow -H_a$ . This introduces the Minkowski metric  
 $\eta_{ab}$, which ensures that we have the
Lorentz group SO(1,10)  rather than the group SO(11). We could also  
have worked with the Cartan involution of equation (2.7) but then Wick 
rotated to Minkowksi signature at any stage. 

The commutators between the generators of K($E_{11}$)Ê are given byÊ[2] 
$$
Ê[J^{ab},J^{cd}] =- \eta^{bd} J^{ac} - \eta^{ac} J^{bd} 
+ \eta^{bc} J^{ad} + \eta^{ad} J^{bc} 
$$
$$
Ê[S^{a_1 \dots a_3}, S_{b_1 \dots b_3}]
Ê = 2S^{a_1\dots a_3}{}_{a_4 \dots a_6}-
18\delta^{[a_1 a_2}_{[b_1 b_2}J^{a_3]}{}_{b_3]} 
$$
$$
Ê[S_{a_1 \dots a_3}, S^{b_1 \dots b_6} ]
Ê = -3S^{b_1 \dots b_6}{}_{[a_1 a_2, a_3]} 
- {5! \over 2} \delta^{[b_1 \dots b_3}_{[a_1 \dots a_3]} S^{b_4 \dots
b_6]}   Ê
\eqno(2.11)$$
The first generator on the right-hand side of the last equation is the
Chevalley invariant combination of the level three and minus three 
generators that we include for completeness,  although it  is beyond the
level truncation used in this paper. 
\medskip
 {\bf 3. The action of K($E_{11}$) on the bosonic fields}
\medskip
In this section we calculate the transformations of the bosonic fields
under rigid K($E_{11}$).Ê By definition, the group element from which the
nonlinear realisation is constructed transforms under rigid
transformations as
$$
Êg \rightarrow g_0 g ,\quad \quad g \rightarrow  g h
\eqno(3.1)$$
where $g_0 \in E_{11}$ is a rigid, i.e. constant,Ê transformation, but $h
\in K(E_{11})$ is a local transformation.Ê
\par
Using the analogue of the Iwasawa decomposition we may write the general
group element of $E_{11}$ asÊ
$$
Êg =  e^ {h^a H_a}  e^{\sum_\alpha A^\alpha E_\alpha} 
e^{\sum_\beta B^\beta S_\beta}
\eqno(3.2)$$
Where the sums over $\alpha$ and $\beta$ run over all positive roots, and
$S_\beta$  denotes an element of $K(E_{11})$. This group element is of
the form of an element of the Borel subalgebra multiplied by aÊ Cartan
involution invariant group element. Using the local symmetry we can
choose the group element to be of the formÊ
$$
Êg =  e^{ h_a{}^b K^a{}_b }  e^{ A_{a_1 \dots a_3} R^{a_1 \dots a_3} } 
 e^{ A_{a_1 \dots a_6} R^{a_1 \dots a_6}} \ldots 
\eqno(3.3)$$
Thus we choose our coset representatives.ÊWe note that we did not use the
local Lorentz group part of $K(E_{11})$ to choose the $h_a{}^b$ to be
symmetric. 
\par
Carrying out a rigid Borel transformation takes us from one coset 
representative to another, so we can immediately read off the
transformation of the fields. However, this is not the case for a general
$E_{11}$ transformation and one has to perform an additionalÊ
compensating local $K(E_{11})$ transformation to bring the group element
back to being one of the coset representatives.
\par
We must also include in the group element a part associated with 
space-time, that is a factor $e^{x^aP_a}$. In principal we should add
further generators associated to the generalised space-time introduced in
[2] corresponding to the non-linear realisation of $E_{11}\otimes_s
l_1$, but these are likely to lead to higher order effects than those
being considered in this paper. As such we take all $E_{11}$ generators
except those of  GL(11) to commute with $P_a$.
\par
We now consider the rigidÊ transformation
$$
Êg_0 =  e^{c^{a_1 \dots a_3} S_{a_1 \dots a_3}}
\eqno(3.4)$$
All other $K(E_{11})$ transformations can be found from this
one by taking commutators.Ê As a result we find at lowest order in the
transformation parameter
$$ \matrix {
Ê\delta_{c^3} e_\mu{}^a
Ê &=& 2c^{\nu\rho\lambda} \left( 9A_{\mu\nu\rho} e_\lambda{}^a
- A_{\nu\rho\lambda} e_\mu{}^a \right) \cr
Ê\delta_{c^3} A_{a_1 \dots a_3}
Ê &=&
ÊÊ -3A_{a_2 a_3 b} \left( 9 c^{\mu\nu\rho} e_\mu{}^c e_\nu{}^d e_\rho{}^b A_{cd a_1}
ÊÊ - c^{\mu\nu\rho} e_\mu{}^c e_\nu{}^d e_\rho{}^e A_{cde} \delta^b_{a_1} \right) \cr
Ê&& +60A_{a_1 \dots a_3 cde} c^{\mu\nu\rho} e_\mu{}^c e_\nu{}^d e_
\rho{}^e
+ c_{\mu\nu\rho} e_{a_1}{}^\mu e_{a_2}{}^\nu e_{a_3}{}^\rho
- c^{\mu\nu\rho} e_{\mu a_1} e_{\nu a_2} e_{\rho a_3}}
 \eqno(3.5)$$
\par
The object $e_\mu{}^a$ is the vielbein and how it enters into the
non-linear realisation is discussed in appendix A. The quantity
$c^3\equiv c^{a_1a_2a_3}$ is the same constant regardless of whether it
carries flat or curved indices. In other words we do not use the vielbein
to convert the flat indices to the curved indices on $c^3$,  but rather
show explicitly the vielbein factors that are present. In other words,
$c^{\mu\nu\rho}=
\delta^\mu_a\delta ^\nu_b\delta ^\rho_c c^{abc} $ is also a constant.
\par
To find the above result one must first move the $g_0$ of equation (3.4)Ê
past the $e^{h\cdot k}$ factor in the group element $g$ of  equation
(3.3) using the equationÊ
$$ 
g_0 e^{h\cdot k} = e^{h\cdot k} e^{-h\cdot k} 
 e^{c^{a_1 a_2 a_3} S_{a_1 a_2 a_3}Ê} e^{h\cdot k} 
$$
$$
= e^{h\cdot k} \exp \{ c_{\mu_1 \mu_2 \mu_3} e_{a_1}{}^{\mu_1}
e_{a_2}{}^{\mu_2} 
 e_{a_3}{}^{\mu_3} R^{a_1 a_2 a_3}  - c^{\mu_1 \mu_2 \mu_3}
e_{\mu_1}{}^{a_1} e_{\mu_2}{}^{a_2} e_{\mu_3}{}^{a_3} R_{a_1 a_2 a_3}  \} 
\eqno(3.6)$$
The presence of the vielbeins in equation (3.6) is explained in the 
appendix. Moving the expression in equation (3.6) after the $e^{h\cdot
k}$ factor past the next
factor in the group element $g$ ,   namely $e^{A_{a_1a_2a_3}
R^{a_1a_2a_3}}$  the
$e^{c^{\mu_1
\dots \mu_3} e_{\mu_1}{}^{a_1} e_{\mu_2}{}^{a_2}e_{\mu_3}{}^{a_3}R_{a_1
\dots a_3}}$ term creates a GL(11) transformation that must beÊ reordered
in the group element. Similar considerations apply to the  passage ofÊ
$e^{c^{\mu_1 \dots
\mu_3} e_{\mu_1}{}^{a_1} e_{\mu_2}{}^{a_2}e_{\mu_3}{}^{a_3}R_{a_1 \dots
a_3}}$ past the factor containing the six form field. Finally, one can
recogniseÊ $e^{c^{\mu_1 \dots \mu_3} e_{\mu_1}{}^{a_1}
e_{\mu_2}{}^{a_2}e_{\mu_3}{}^{a_3}R_{a_1 \dots a_3}}$ as part of the
compensating local transformationÊ
$$
h=e^{c^{\mu\nu\rho}e_\mu{}^a e_\nu{}^b e_\rho{}^c S_{abc}}
\eqno(3.7)$$
We note that this contains a term $e^{c^{\mu\nu\rho}e_\mu{}_a e_\nu{}_b
e_\rho{}_c R^{abc}}$ which must be reabsorbed into the change in the
three form field together with the similar term that arises from the
passage of the factor $e^{ c_{\mu_1 \mu_2 \mu_3} e_{a_1}{}^{\mu_1}
e_{a_2}{}^{\mu_2} 
 e_{a_3}{}^{\mu_3} R^{a_1 a_2 a_3}}$  in equation (3.6). 

To calculate the variation of the vielbein under $S_6$ we repeat this
procedure with $g_0 =  e^{c_{a_1 \dots a_6} S^{a_1 \dots a_6}} $ and a
suitably chosen compensating local transformation.Ê We find the result
$$
Ê\delta_{c^6} e_\mu{}^a =
Ê 5! c^{\nu_1 \dots \nu_6} A_{\nu_1 \dots \nu_6} e_\mu{}^a
Ê - 5! 9 c^{\nu_1 \dots \nu_6} A_{\nu_1 \dots \nu_5 \mu} e_{\nu_6}{}^a
Ê - {5!9 \over 2} c^{\nu_1 \dots \nu_6} A_{\nu_1 \dots \nu_3} A_{\nu_4 \nu_5 \mu} e_{\nu_6}{}^a
\eqno(3.8)$$
\par
Finally we write down the effect of a rigid Lorentz transformation on the
vielbein in this formalism so as to fix the normalisation. That is we
take 
$g_0=e^{ c_{ab} J^{ab}}$ and process it as in equation (3.6) to find a
  local transformation.Ê The result is 
$$
Ê \delta _{c^2}e_\mu{}^a = 2 c_\mu{}^\nu
e_\nu{}^a
\eqno(3.9)$$
where the second index on $ c_\mu{}^\nu$ is simply now written as an upper
index. 
\medskip
{\bf 4. Spinorial representations of K($E_{11}$)}
\medskip
In this paperÊ we wish to include fermions in the $E_{11}$
non-linear realisation. As we have already mentioned the  prototypical
example is theÊ maximal supergravity in four
dimensions which has an
$E_7$ symmetry [6]. In this theory, and indeed all supergravity theories
in which the scalars are part of the supergravity multiplet,   
theÊ spinors appear in the nonlinear realisation 
 as matter representations. The matter representationsÊ transform as
a linear representation of the chosen local subalgebra, which is SU(8) in
the example just considered. This isÊthe Cartan involution invariant
subalgebra and so  theÊ maximal compact subgroup of $E_7$. We note that
once one has chosen a coset representative, one must in general carry out
compensating local transformations, which act on matter representations.Ê

Spinors have already been introduced in the $E_{10}$  approach [8-10]
where they also took the spinors to transform under the Cartan involution
invariant subalgebra.Ê To construct the representation of K($E_{10}$)
appropriate to the gravitino, these authorsÊ started with the vector
spinorÊ representation of SO(10)Ê and introduced a transformationÊ for
$S_3$, up to level three, thatÊ satisfied the known commutation relations
for the $K(E_{10})$.Ê It turned out that it was enough at low levels to
introduce only the gravitino field and so the representation found was
highly unfaithful.Ê
\par
These techniques also apply to  $E_{11}$ and we also take the gravitino
to be a matter representation. We start with  the standard Lorentz
transformation of the gravitino SO(10,1) with a tangent space vector
index; 
$$
ÊJ_{ab} \psi_c =- {1 \over 2} \gamma_{ab} \psi_c - 2 \eta_{c[a}\psi_{b]}
\eqno(4.1)$$
To find a suitable transformation of the vector spinor under $S_3$ we
write down all possible terms with the correctÊ SO(1,10) character and
demand that it obey the algebra given in equation (2.11) involving the
$S_3$ generator. In particular from the commutator between two $S^3$
generators in equation (2.11), one derives the following two relations
$$
Ê[S^{abc},S_{ade}] = 0,\quad Ê
Ê[S^{abc},S_{abd}] = - J^c{}_d Ê
\eqno(4.2)$$
Where $a,b,c,d,e$ are distinct indices.The second relation relates the
$S_3$ transformation back to the known SO(1,10) transformation of
equation (4.1). Given the  $S_3$ transformation we can find  all
higher level $K(E_{11})$ transformations by taking repeated commutators. Ê
We find that the transformations of the vector spinor, that is the
gravitino,  up to level two, are given byÊ
$$ \matrix {
Ê&&S_{abc} \psi_d =
Ê {1 \over 2} \gamma_{abc} \psi_d 
-  \gamma_{d [ab} \psi_{c]} + 4  \eta_{d [a} \gamma_b \psi_{c]} \cr
Ê&&S_{abcdef} \psi_g =
Ê -{1Ê \over 4} \gamma_{abcdef}\psi_g - 2 \gamma_{g[abcde} \psi_{f]} 
+ 5 \eta_{g[a}\gamma_{bcde}\psi_{f]} \cr
}Ê \eqno(4.3)$$
\par
One can repeat this procedure, starting with the spin 1/2 representation
of SO(10,1) and recover the result [2]Ê
$$
ÊJ_{ab} \psi = -{1 \over 2} \gamma_{ab} \psi \quad 
ÊS_{abc} \psi = {1 \over 2} \gamma_{abc} \psi \quad
ÊS_{abcdef} \psi = -{1Ê \over 4} \gamma_{abcdef}\psi 
 \eqno(4.4)$$

\medskip
{\bf 5. Commutator of K($E_{11}$) and Supersymmetry}
\medskip

In this section we will calculate the commutator of the supersymmetry
variations  and the $K(E_{11})$ transformations on the vielbein and  the
three form. For our supersymmetry variations we take the well known
transformations from eleven dimensional supergravity. We will find that
the commutators result in symmetries of the theory and so demonstrate the
consistency of $E_{11}$ with supersymmetry at least at low levels. This
is far from guaranteed as $E_{11}$ has so far been based entirely on the
bosonic fields.Ê Ê We takeÊ the supersymmetry transformations of the
vielbein, the three form, and its dual, the six form with the Grassmann
parameter $\epsilon_\alpha$ to be [13] 
$$ \matrix {
Ê\delta_{\epsilon} e_\mu{}^a &=& \bar{\epsilon} 
\gamma^a \psi_\mu \cr
Ê\delta_{\epsilon} A_{\mu\nu\rho} 
&=& {1 \over 2} \bar{\epsilon} \gamma_{[\mu\nu}\psi_{\rho]} \cr
Ê\delta_{\epsilon} A_{\mu_1 \dots \mu_6} &
=& -{1 \over 60} \bar \epsilon \gamma_{[\mu_1 \dots \mu_5} \psi_{\mu_6]}Ê
+{1 \over 2} \bar \epsilon \gamma_{[\mu_1\mu_2} \psi_{\mu_3} A_{\mu_4
\dots \mu_6]} \cr }Ê
\eqno(5.1)$$
We note that the normalisation of the fields was already
determined by their appearance in the $E_{11}$ group element of equation
(3.3) and those chosen in equation (5.1) are the ones compatible with
this previous choice. 

One finds that the commutator of the variation of  $Q_\alpha$ and $S_3$ on
the
vielbein is given byÊ
$$
Ê[\bar \epsilon Q, c^3 \cdot S_3] e_\mu{}^a =
ÊÊ {1 \over 2} c^{bcd} \bar {\epsilon} \gamma_{bcd} \gamma^a \psi_\muÊ
Ê Ê - 4  c_\mu{}^{ad} \bar{\epsilon} \psi_dÊ
ÊÊ - 4  c_\mu{}^{cd} \bar{\epsilon} \gamma^a{}_c \psi_d
ÊÊ + 4  c^{acd} \bar{\epsilon} \gamma_{\mu c} \psi_d
ÊÊ +Ê  c^{bcd} \bar{\epsilon} \gamma^{a}{}_{\mu bc} \psi_dÊ
\eqno(5.2) $$
When carrying out this calculation it is important to remember that the
K($E_{11}$) transformation of the gravitino discussed in section three
was defined in the tangent frame, however the gravitino in the
supersymmetry transformations has a curved index, so when considering the
K($E_{11}$) variation of the gravitino, we must include the vielbein
required to convert a flat to a curved index, that is
$\psi_\mu = e_\mu{}^a
\psi_a$. The same applies toÊ the threeform which we must write as
$A_{\mu\nu\rho} = e_\mu{}^a e_\nu{}^b e_\rho{}^c A_{abc}$.

From equation (5.2) we extract the generic form of the commutatorÊ
$$
Ê[Q, S_{bcd}] =
Ê {1 \over 2}\gamma_{abc} Q +  \left(
Ê Ê {1\over  2} \gamma^f{}_{ebc}
Ê Ê +2 \eta_{eb} \delta^f_c
Ê Ê - 2 ( \eta_{eb} \gamma^f{}_c - \delta^f_b \gamma_{ec})Ê
Ê \right) \psi_d J_L{}^e{}_f
\eqno(5.3)$$
which we recognise as a supersymmetry transformation and a {\bf local } 
Lorentz transformation denoted by the symbol $J_L{}^e{}_f$.  On the
metric, which is a Lorentz  invariant object, the field dependent Lorentz
transformations do not appear, and we are left with
$[Q, S_{abc}] = {1 \over 2}\gamma_{abc} Q$. This is expected, because the
supercharge
$Q$ is a spinor, which transforms as in equation (4.4). 

We note that the commutator (5.3) is field dependent. This is a well known
phenomenon that occurs in the commutator of supersymmetry and gauge
transformations when some of the fields have been set to zero using the
supermultiplet of gauge symmetries, the prototype example is  to fix 
the Wess-Zumino gauge in supersymmetric Yang-Mills theory; for a
review see [14]. It is to be expected here as we have used a local
symmetry, that is the
$K(E_{11})$, to gauge away the non-Borel part of the group element.
\par
A similarÊ calculation on the threeform field gives
$$
Ê{[\bar \epsilon Q, c^3 S_3]} A_{\mu\nu\rho} 
= {1 \over 4} c^{abc} \bar \epsilon
\gamma_{abc} \gamma_{\mu\nu} \psi_\rho
\eqno(5.4)$$ 
The spacetime threeform is a Lorentz invariant object, and one does not
expect to see the field dependent terms of equation (5.3). Thus one finds
that the generic commutator of a supersymmetry transformation with a rigid
$S^{a_1a_2a_3}$ transformation is, up to level two,  of the form 
$$
Ê[Q, S_{abc}] = {1 \over 2}\gamma_{abc} QÊ
\eqno(5.5)$$
plus local transformations.
\par
The commutator of supersymmetry and $S_6$ on the vielbein is given byÊ
$$ \matrix{
Ê{[\bar \epsilon Q, c^6 S_6]} e_\mu{}^a =
Ê &+{1 \over4} c^{\nu_1 \dots \nu_6} 
\bar \epsilon \gamma_{\nu_1 \dots \nu_6} \gamma^a \psi_\mu \cr
Ê & - {5! \over 2} c^{\nu_1 \dots \nu_6} 
\bar \epsilon \gamma_{\nu_1 \nu_2} \psi_{\nu_3} (9 A_{\mu\nu_4 \nu_5}
e_{\nu_6}{}^a - A_{\nu_4 \dots \nu_6}e_\mu{}^a) \cr Ê&+2 c^{\nu_1 \dots
\nu_6}\bar \epsilon \gamma^a{}_{\mu\nu_1 \dots \nu_5} \psi_{\nu_6}Ê ÊÊ
-20 c_\mu{}^{a\nu_1 \dots \nu_4} \bar \epsilon \gamma_{\nu_1 \dots \nu_3}
\psi_{\nu_4} \cr ÊÊ &-5Ê \left( ÊÊ c_\mu{}^{\nu_1 \dots \nu_5} \bar
\epsilon \gamma^a{}_{\nu_1 \dots \nu_4} \psi_{\nu_5} ÊÊ -c^{a\nu_1 \dots
\nu_5} \bar \epsilon \gamma_{\mu \nu_1 \dots \nu_4} \psi_{\nu_5} Ê \right)
Ê} \eqno(5.6)Ê $$
These variations lead to the commutator relation 
$$ \matrix {
 [Q, S_{a \dots f}] =&
  {1 \over 4} \gamma_{a \dots f} Q
  - 30 \gamma_{ab} \psi_{c} S_{def }  \cr
  &+ \left(
   {5 \over 2} ( \delta_{f j} \gamma^k{}_{a \dots d}
\psi_{e} - \delta^k_{f} \gamma_{j a \dots d} \psi_{e}
)
  + \gamma^k{}_{j a \dots e} \psi_{f}
  - 10 \delta_{e j} \delta^b_{f} \gamma_{a \dots c}
\psi_{d}
  \right) \psi_{f} J_L{}^j{}_k
}\eqno(5.7)$$where the right hand side is understood to be antisymmetrised over the
indices $abcdef$. Thus we may write the commutator as 
$$
Ê[Q, S_{abcdef}] ={1 \over 4}\gamma_{abcdef}Q
\eqno(5.8)$$
plus local transformations. We note that equations (5.5) and (5.8) are
compatible with regarding the supercharge as a spinor which we found to
transform as in equation (4.4). The commutators of the Cartan involution
subalgebra with the supersymmetry  as anticipated in [2].  

 \medskip
{\bf Appendix A. Vielbeins in $E_{11}$}
\medskip
In the calculations given in this paper the vielbein plays an
important role, and in this appendix we briefly discuss how the vielbein
appears in the $E_{11}$ non-linear realisation. For this purpose we
can take  our group elementÊto contain just the part appropriate for
gravity, namely 
$$
Êg=e^{x^aP_a} e^{h_a{}^b K^a{}_b}\ldotsÊ
\eqno(A.1)$$
where $\ldots$ indicates factors involving higher level fields.Ê
The most direct way to see the presence of the vielbein is to 
compute the Cartan formÊ
$$
Ê{\cal V}= g^{-1} \partial_\mu g
= dx^\mu e_\mu{}^a  (P_a+(e^{-1}\partial_\mu e)_a {}^b K^a{}_b+ \ldotsÊ)
\eqno(A.2)$$
where $e_\mu{}^a= (e^h)_\mu{}^a$. The Cartan forms transform under the
local subalgebra $K(E_{11})$ as ${\cal V}\to h^{-1} {\cal V} h + h^{-1}
d h$. AtÊ lowest level this is just the Lorentz group and so
$e_\mu{}^a$ transforms on its upper $a$ index just like a vector under
the Lorentz group while any reparameterisation, more precisely any GL(11)
transformation,  of
$x^\mu$ gives a corresponding change in Ê the lower $\mu$ index of
$e_\mu{}^a$. Thus
$e_\mu{}^a$Ê does transform as a vielbein should. Indeed constructing the
theory of gravity from the non-linear realisation as was first done in
[15], and again in a more vielbein orientated approach in [16], one finds
that $e_\mu{}^a$ does indeed appear in the theory as the vielbein should.
\par
Effectively, the above calculation of the vielbein  evaluates
$e^{h_a{}^bK^a{}_b}$ in the vector representation as this factor acts on
$P_a$. In this representationÊ
$$
Ê\left( K^a{}_{b} \right)_c{}^d = \delta ^a_{c} \delta_b{}^d
\eqno(A.3)$$
where $c,d$ are the representation matrix indices  clearly giving
$(e^{h_a{}^bK^a{}_b})_c{}^d=e^h_c{}^d$ in vector representation.Ê
\par
In the paper we encounter expressions where we  move $e^{h_a{}^bK^a{}_b}$
past generators in representations of  GL(11), for example equation
(2.4). In particular we find that
$$
Ê e^{-h \cdot K} c^{a_1 \dots a_3} R_{a_1 \dots a_3} 
 e^{h \cdot K} = c^{\mu\nu\rho} e_\mu{}^a e_\nu{}^b e_\rho{}^c R_{abc}
\eqno(A.4)$$
We recall that the parameter $c^3$ is the same constant  no matter what
indices it displays, but it is natural to write its indices so as to
reflect what it is contracted with. We also give the analogous result for
the positive root generatorsÊ
$$
Ê e^{-h \cdot K} c_{a_1 \dots a_3} R^{a_1 \dots a_3} 
 e^{h \cdot K} = c_{\mu\nu\rho} e_a{}^\mu e_b{}^\nu e_c{}^\rho R^{abc}
\eqno(A.5)$$
which involves the inverse vielbeins $e_a{}^\mu 
= (e^{-h \cdot K}) {}_a{}^\mu$.Ê

\medskip
{\bf 8. References}
\medskip
\item{[1]} P. West, {\sl $E_{11}$ and M theory}, Class. Quant. Grav. {\bf
18} (2001) 4443, {\tt hep-th/9501068}
\item{[2]} P.  West, {\sl $E_{11}$, SL(32) and Central Charges} Phys.
Lett. {\sl B575} (2003) 333, {\tt hep-th/0307098}. 
\item{[3]} F. Riccioni, Duncan Steele, Peter West, 
{\sl E11 origin of all gauged supergravities};  JHEP 0707 (2007) 63, 
arXiv:0705.0752.
\item{[4]} E.  A. Bergschoeff, T. Nutma and  I.  De Baetselier, 
{\sl $E_{11}$ and the embedding tensor}, JHEP {\bf 9} (2007) 047, {\tt
arXiv:0705.1304}
\item{[5]} F. Riccioni and P. West, 
{\sl E(11)-extended spacetime and gauged supergravities}, JHEP {\bf
0802} (2008) 039, {\tt arXiv:0712.1795}
\item{[6]} E. Cremmer and  B. Julia, {\sl The SO(8) supergravity}, Nucl.
Phys. {\bf B 159} (1979) 141
\item{[7]} T. Damour, M. Henneaux and  H. Nicolai {\sl E10 and a Small
Tension Expansion of M Theory} Phys. Rev. Lett. {\bf 89} (2002) 221601
{\tt hep-th/0207267}
\item{[8]} S. de Buyl, M. Henneaux and  L. Paulot, {\sl Extended E8
Invariance of 11-Dimensional Supergravity} JHEP {\bf 0602} (2006) 056
{\tt hep-th/05122992}
\item{[9]} T. Damour, A. Kleinschmidt qand  H. Nicolai {\sl
Hidden symmetries and the fermionic sector of eleven-dimensional
supergravity} Phys. Lett. B {\bf 634} (2006) 319 {\tt hep-th/0512163}
\item{[10]} S. de Buyl, M. Henneaux and  L. Paulot {\sl Hidden
Symmetries and Dirac Fermions}Ê Class. Quant. Grav. {\bf 22} (2005) 3595
{\tt hep-th/0506009}
\item{[11]} M Henneaux, E Jamsin, A Kleinschmidt and  D Persson; Phys.
Rev. D (2009) 045008; arXiv:0811.4358
\item{[12]} M. R Gaberdiel, D. I. Olive and  P. West; {\sl class
of Lorentzian Kac-Moody algebras} Nucl. Phys. {\bf B645} (2002) 403 {\tt
hep-th/0205068}
\item{[13]} E. Cremmer, B. Julia, and J. Scherk, 
{\sl Supergravity Theory In Eleven Dimensions} ,Phys. Lett. {\bf B76}
(1978) 409-412; 
Igor Bandos, Nathan Berkovits and Dmitri Sorokin,
{\it Duality-Symmetric Eleven-Dimensional Supergravity and its 
Coupling to M-Branes},
Nuclear Physics {\bf B 522} (1997) 214-233,
{\tt hep-th/9711055}. 
\item{[14]} P. West, {\sl Introduction to Supersymmetry and
Supergravity}, World ScientificÊ (1990)
\item{[15]} A. Borisov and V. Ogievetski, {\sl Theory of dynamical affine
and conformal symmetries as the theory of the gravitational field}, Teor.
Mat. Fiz. {\bf 21} (1974) 329
\item{[16]} P. West, {\sl Hidden Superconformal Symmetry in M Theory},
JHEP {\bf 08} (2000) 007, {\tt hep-th/005270}

\end